\documentclass[preprint,12pt]{elsarticle}
\usepackage{xcolor}
\usepackage{float}
\usepackage{graphicx}
\usepackage{amssymb}

\usepackage{multirow}

\journal{}

\bigskip
\begin{document}

\begin{frontmatter}


\title{An affordable tool based on a pedestrian-vehicle collision model to support the  fieldwork and reconstruction}



\author[1]{Ezequiel Mart\'inez}
\author[2]{Juan Pablo Vargas}
\author[3]{Alejandra Baena}

\address[1]{Defensor\'ia del Pueblo de Colombia, Forensic Physics Group, Bogot\'a DC, Colombia}
\address[2]{Univision Communications Inc, Bogot\'a DC, Colombia}
\address[3]{Universidad Antonio Nari\~no, Physics Deparment, Forensic Physics Line Research, Bogot\'a DC, Colombia}

\begin{abstract}
A Free access tool based on a pedestrian-vehicle collision model is presented. The model permitted the qualitative and quantitative description of the event's whole dynamic by segments called pre-collision, collision, and post-collision. Furthermore, it enabled the determination of the magnitude of the vehicle's initial speed before the collision with a pedestrian and the location of the point position of impact on the road where the accident occurred. The model's inputs are related to the evidence collected at the scene, which provides a checklist platform for supporting investigators' fieldwork. Additionally, the pre-collision segment permitted the investigators to develop an avoidability study that may contribute to road safety evaluation.
The model was validated, comparing the results statistically with experimental cases developed with dummies, bodies, and reconstructed cases. It is shown that there is no significant difference, thus verifying its functionality. The tool is available as a mobile app in Spanish and English, allowing significant affordability to investigators from some low and middle-income countries.
\end{abstract}


\end{frontmatter}


\section{Introduction}
\label{S:1}
According to the WHO \cite{WHO}, about 1.35 million people die per year due to a traffic accident. Between 20 and 50 million people suffer no fatal injuries; however, these may incur disabilities and some affection for their health. Regarding the age at risk, the reports present that road traffic injuries are the leading cause of death for children and young adults among 5-29 years. Considering the socioeconomic status, more than 90\% of deaths occur in low and middle-income countries.
\\
\\
On the other hand, it is crucial to consider the risk related to road users. The more vulnerable are pedestrians, cyclists, and motorcyclists, and they constitute more than half of the road traffic deaths about $54\%$. In Europe, the roads are safer globally; however, 21  \%  of the whole traffic fatalities are pedestrian \cite{EC}. The number of deaths is a severe problem in other regions as South East Asia, where vulnerable road users constitute almost 75  \%  of accident traffic fatalities. In the regions of the Americas \cite{paho}, pedestrians, cyclists, and motorcyclists make up 45 \% of road traffic deaths.
\\
\\
Therefore, road safety is considered a  public health problem that extends to an international level. It has been included as a priority goal in the Sustainable Development Agenda for 2030 determined in 2015 by all member states of the United Nations General Assembly (UNGA)\cite{paho}. All of the strategies proposed a particular intervention for every country that contemplates the causes of accidents in each context to establish new policies to reduce the road risk and, consequently, the traffic accident numbers \cite{savelives}. 
\\
\\
To determine the main risk factors that generate traffic accident is crucial to study, analyze, investigate and represent the accident graphically.  This technical and scientific procedure is well known as a traffic collision reconstruction \cite{Rivers2}. It permits to obtain conclusions about the causes of the accident and the facts during a traffic collision. In the reconstruction process, physics and engineering principles are applied, supported sometimes by software and simulation to facilitate the calculations and visual representation \cite{Lozano1998213}. 
\\
\\
The reconstruction procedures are mainly developed by experts such as forensic engineers, specialized units in law enforcement agencies, or private consultants. Subsequently, the results may provide substantial impacts in solving judicial cases, make roads \cite{Bobermin2021} and motor vehicle aspects safer \cite{Yuan2017370}, and improve the pedagogic strategies focusing on the best practices and the human behavior on the roads \cite{Hou20211}.
\\
Unfortunately, the appropriate study of traffic accidents is not achieved in many cases. For instance, many low and middle-income countries present some difficulties in developing acceptable investigations due to many factors. One of them is the inability to access costly software to support the reconstructions; most specialized training is in foreign languages, inconvenient \cite{baena20,remolina20}. Another factor is that law enforcement doesn't have experts with the knowledge necessary in physics and engineering to face accident study optimally in many cases. It happens in most unprivileged urban cities or isolated regions \cite{su11226249}. 
\\
\\
Furthermore, the study of traffic accidents involving vulnerable actors as the pedestrian is a challenge in many particular cases. The available models, the collection, and the evidence's interpretation at the scene could be insufficient for a proper reconstruction \cite{Dario}. For modeling, it is crucial to establish the accident particularities to get a model for each accident. Another way is applying general models that share physical features with the case of the study—being this the usual procedure to reconstruct an accident in many places \cite{Daily08}.
\\
\\
One of the most applicable models to traffic accident reconstruction is the frontal collision vehicle-pedestrian. It depends on variables as the vehicle geometry, height of pedestrian, masses, and environmental factors. It allows knowing kinematic magnitudes as the vehicle's initial speed before the collision,   pedestrian launch speed, and impact zone \cite{Raymond10,Lopez04}. Some of them also permit the analysis of damage of cars, pedestrian injuries, and pedestrian launch distance \cite{Toor2002}. 
\\
\\
Although the estimations given by the models are without a doubt very useful for the traffic accident investigation goal, there are some aspects not considering yet as the physics at the pre-collision moment. It analysis may give information about the avoidability of accidents and other substantial details, which could improve the road safety policies and the knowledge of human road behavior. Given the above, in this paper, a handly model for vehicle-pedestrian collision is presented. This model includes analyzing various events into the dynamical of this kind of accident,  constituted by pre-collision, collision, and post-collision segments, which incorporate phases of the process related to either the pedestrian and the vehicle.
\\
\\
The model presented is the input for a free mobile app called ARTgrama, currently available in Spanish and English. This mobile app permits including and checking the information collected at the accident scene during the fieldwork and its reconstruction. 
Likewise could be used to investigate the accidents that involve other vulnerable actors as cyclists and motorcyclist. This mobile App contributes to assisting law enforcement, traffic collision experts, and academic purposes, especially from the low and middle-income countries, where the training and software are inaccessible in many cases.
\\ 
This paper was organized as described following: Section \ref{MD} describes the collision vehicle-pedestrian mechanism for the modeling. Section \ref{model} shows the physical principles and equations used to build the model. Later, Section \ref{results} shows the model validation method comparing the results with some experimental cases previously studied. The conclusions and discussion are presented in Section 5.

\section{Mechanism Description of Collision Vehicle-Pedestrian}
\label{MD}
The collision process considers a series of events during the accident. It is presented by moments represented by three segments, called pre-collision, collision, and post-collision. Each moment is explained as follows:

\subsection{Pre-Collision}
This segment starts with stage I, where the driver perceives the pedestrian on the road (point of perception $P_{p}$), continuing for the decision and ending with the reaction point $P_{r}$. The total time of this stage is called perception-decision-reaction time $t_{pdr}$ at a distance $d_{pdr}$, and it is about 0.4s-0.75s \cite{Asthon89,Brown02}. Here, it is assumed the speed is a constant and the acceleration is zero.
\\
After the point, $P_{r}$, starts stage II  or the reaction which is related to the braking process, where the foot's movement from the accelerator to the pedal braking is considered, and its time is $t_{a-f}$ that is about to 0.2s-0.3s. The braking system generates a deceleration of the vehicle until the wheels lock up; this mechanical response time interval is  $t_{rm}$ = 0.4s-0.75s. Therefore, the total reaction time $t_{rt}$ includes the summation of $t_{pdr}$, $t_{a-f}$ and $t_{rm}$ which is about 1.2s-1.5s  in normal conditions \cite{Brown02,Brach10}. Since the brake is applied, the tires' angular and tangential speed decreases, which increases the friction force and coefficient up to a maximum peak. Hence, it begins with a weak locking of the tires where the marks on the road are faint (shadows); then, the top lock of the tires is reached at maximum (zero angular speed) \cite{Reed89}, and then the marks become intense until the stopping of the vehicle. The tire marks' length gives the distance  $d_{hf}$ of this reaction stage. The acceleration is assumed as a constant when the braking marks are intense on the road, namely, from the beginning of the tire lock. Here. Therefore the kinetic friction coefficient ($\mu_v$) could be estimated using tables from previous experiments or with a braking test with vehicles at the location for more precision \cite{SAE10,Walter83}. The braking may happen at this segment (pre-collision) or after the interaction between vehicle-pedestrian (collision segment).
\\
\subsection{Collision}
After the pre-collision, the collision segment continues. It may be with one or two impacts between the vehicle and pedestrian, and the interaction is for a time and distance named here as ($t_{cv}$) and ($d_{cv}$). Here, the vehicle and pedestrian impact may be assumed as an inelastic collision; therefore, the restitution coefficient equals zero ($e=0$). After this stage, the  movement of the car slows down, and it is possible to observe the registration of tire marks on the road.
\\
When the double impact happens, it usually occurs for low frontal geometry vehicles; the time of interaction is about $t_{cp}$ = 0.1s \cite{Van83,Kall89}, at a distance ($d_{cp}$). The vehicle hits the pedestrian's lower extremities with its bumper, blinds, streetlights; it causes damage to the vehicle and injuries to the pedestrian. The impact occurs under the center of gravity of the pedestrian, transmitting linear and angular moments, which makes the pedestrian get up while moving forward, later rotating and wrapping the front of the vehicle, sliding on the engine cover (damage ), and receiving a second impact with the panoramic (damage).  It causes injuries to the head, trunk, and upper extremities of the pedestrian projected forward relative to the vehicle's movement with an inclination angle ($5^\circ \leq \theta \leq 24^\circ$) \cite{Ezequiel14}. Likewise, the pedestrian flights a distance ($d_{v}$) and collides with the road surface, then rotates, jumps, flips, and crawls in a sliding length ($d_{d}$) and reaches its final resting position. Here the drag coefficient of friction pedestrian/road (asphalt and dry) could be assumed $0.7 \leq \mu_{p} \leq 0.8$ \cite{Brach10,Wood00,Hill94}. The evidence of the dragging is visible on the road.
\\
When one impact happens, it usually occurs for high frontal geometry vehicles. The interaction time is about $t_{c}$ = 0s  at a distance ($d_{c}$=0m) \cite{Brach10,Ezequiel14}. For instance, a bus hits the pedestrian's lower and higher extremities with its bumper, blinds, and streetlights. This impact is above the center of gravity of the pedestrian. It is projected forward, initiating a flight movement that corresponds to a flight distance ($d_{v}$). Then, it collides with the road, rotates and slides on it, and reaches its final resting position in a sliding distance ($d_{d}$) with a kinetic friction or drag coefficient of the pedestrian/road ($\mu_p$).  
\\
It is important to remark that it is possible to observe marks of tires, fragments, and traces related to the collision, indicating the exact point of the impact vehicle-pedestrian. Nonetheless, some factors don't allow the registration of this evidence for all cases.  The environmental conditions (wet, puddles of water, mud, inadequate lighting at night) or lack of information do not permit the police to identify the specific location due to the absence of witnesses, cameras, and others. However, the vehicle's final position and the pedestrian or the vehicle and the pedestrian's blood pool are usually presented in the sketch of the polices report. It is crucial information that permits estimating the distance thrown, the impact point, and vehicle speed before the collision.
\\

\subsection{Post-Collision Stage}
In the post-collision stage, there is a detachment of either the vehicle and the pedestrian, it corresponds to an inelastic collision, and after the impact, they have the same speed, $V_{vd}=\alpha V_{pd}$ ($\alpha=1$). Later, they continue with an independent trajectory with constant deceleration. It process may or not leave marks until its final position ($d_{vd}$).

\subsection{Accident avoidability}
In the pre-collision moment, the distance of the tire marks on the road may determine the magnitude of the vehicle's initial speed at the collision segment. However, it depends on if the tires blocks before or after the impact point. In this case, the initial velocity is called $V_{va}$. Once the speed magnitude before the collision $V_{va}$ is estimated, it enables the study of the possibility of avoiding the accident is developed. It means the analysis from the point of perception and the point of the driver's reaction.
\\
\\
The accident process explained before by segments established that the pedestrian and the vehicle have interdependent movements. Therefore, kinematic physical variables (distance, time, speed, and acceleration) and dynamics ( moment, restitution, and mass ratio)  allowed the modeling for pre-collision, collision, and post-collision stages.

\section{Modeling of Collision Vehicle-Pedestrian}
\label{model}

\begin{figure}[htp]
\begin{center}
\includegraphics[scale=2.5]{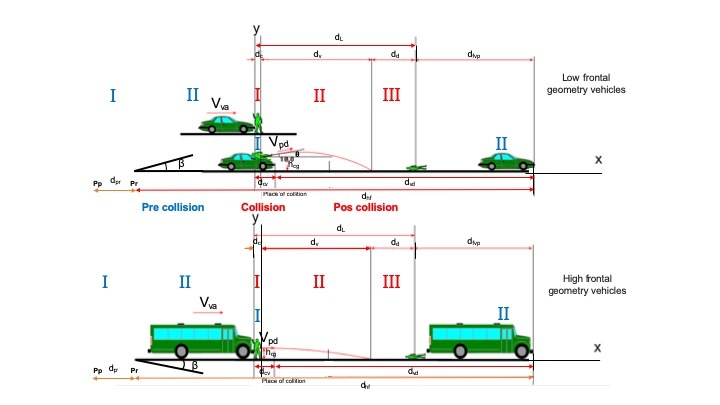}
\caption{Pre-collision, collision, and post-collision segments for a vehicle-pedestrian accident are shown. Stages on blue are related to the vehicle and the red ones for the pedestrian.}
\label{col}
\end{center}
\end{figure}

Fig.\ref{col} shows the process of the accident with all variables considering for modeling. For the pre-collision segment, stages I and II  (on blue) are related to perception and reaction, respectively; these are both associated with the vehicle's driver. Two interdependent movements are assumed for collision and post-collision segments, one for each road user (vehicle and pedestrian). For the pedestrian, stages I, II, and (in red) are related to contact, flight, sliding. On the other hand, stages I and II (on blue) are the contact and deceleration for the vehicle.

The main objective of the model is to estimate the initial speed of the vehicle at the collision segment as a function of the launch distance $V_{va}=f(d_{L})$. It is estimated considering the cinematic and dynamic variables as launch angle $\theta$, the height of the pedestrian's center of gravity ($h_{cg}$), the angle of the tilt of the road $\beta$, the drag coefficient of the pedestrian/road $\mu_{p}$ and the friction coefficient of the vehicle/road. The variables values must be obtained empirically on the field investigation or from previous research experiments at the same conditions as the accident.
\\
It starts the modeling by contemplating some assumptions supported by the mechanism of the accident described above in section \ref{MD}. 
The primary impact point of the pedestrian with the vehicle to the final position on the road is defined as the  pedestrian launch distance ($d_{L}$) which is the summation of the contact distance ($d_{c}$), the flight distance ($d_{v}$) and the sliding distance ($d_{d}$) as in Eq. \ref{dis}. $d_{c}$, $d_{v}$, $d_{d}$ are distances determine for the pedestrian and it belong to Stage I, II and III (on red) respectively on Fig. \ref{col}.

\begin{equation}
    d_{L}=d_{c}+d_{v}+d_{d}
\label{dis}
\end{equation}

According to previous reports \cite{In11}, the contact distance of the pedestrian  $d_{c}$ may be estimated as the half of the initial speed of the launch for pedestrian $V_{pd}$  by the contact time of the pedestrian with the vehicle $t_{c}$, as follows in Eq \ref{dc}:

\begin{equation}
    d_{c}=\frac{1}{2}V_{pd}t_{c}
\label{dc}
\end{equation}
The ratio $\alpha$ between the pedestrian initial launch speed ($V_{pd}$) and the speed of the vehicle at contact ($V_{vd}$) is 1, thus $V_{pd} =V_{vd}$. On the other hand, from the initial collision until the launch, the pedestrian reaches the magnitude of the vehicle speed; as a consequence, it decreases its momentum, expressed by Eq \ref{Vva}. 

\begin{equation}
V_{va} = (m_{v}+m_{p})V_{vd}/m_{v} 
\label{Vva}
\end{equation}

$m_{v}$ and $m_{p}$ correspond to the mass of vehicle and the mass of pedestrian, respectively.

The flight distance is determined by the parabolic trajectory of the pedestrian center of gravity. Here, the first term is the horizontal component of the flight, which is the time $t_{v}$ by the launch's initial speed $V_{pd}$. The second term is the vertical component of the flight according to uniformly accelerated motion. $\beta$ is the tilt of the road.

\begin{equation}
d_{v} =t_{v} V_{pd} \cos \theta –  t_{v}^2 \frac{1}{2} g \sin \beta
\label{dv}
\end{equation}

The value of the flight time ($t_{v}$) is estimated from the launch position to the place where the  pedestrian impacts the road surface as Eq.\ref{tv}

\begin{equation}
t_{v}=\frac{V_{pd} \cos \theta }{g \cos \beta} + \frac{\sqrt{V_{pd}^2 \sin ^2 \theta  + 2g h_{cg} \cos \beta}}{g \cos \beta}
\label{tv}
\end{equation}

Here,  $\theta$ corresponds to the angle of launch, $\beta$ is the tilt of the road, $g$ the gravity, and $\mu_{p}$ is the pedestrian's drag coefficient.
\\
The final part of the launch distance is the sliding distance of the pedestrian. It could be estimated as in Eq \ref{dd}, according to the kinematics of two-dimensional movement. 
 
 \begin{equation}
 d_{d}= \frac{V_{pd} (\cos \theta + \mu_{p} \sin ^2 \theta)}{2g(\mu_{g} \cos \beta \pm \sin \theta)}
 \label{dd}
\end{equation}

Considering Eqs.\ref{dis} to \ref{dd}, it is possible to calculate the initial speed of the launch for pedestrian as a function of the distance of launch $V_{pd} = f(d_{L})$  as presented in Eq \ref{Vpd}.

\begin{equation}
Vpd=\frac{g(\mu_{p} \cos \beta \pm \sin \beta)}{(\cos \theta + \mu_{p} \sin ^2 \theta)} \left(\sqrt{\frac{t_{c}^2}{4}-\frac{2(\cos \theta + \mu_{p} \sin \theta)^2}{g(\mu_{p} \cos \beta \pm \sin \beta)}(h_{cg}\mu_{p}-h_{cg} \tan \beta - d_{L})}-\frac{t_{c}}{2}\right)
 \label{Vpd}
\end{equation}

Where, $V_{pd}$ is the initial speed of launch for pedestrian, g is the gravity, $\mu_{p}$ corresponds to the drag coefficient, $\theta$ is the launching angle, $\beta$ is the tilt of the road; $h_{cg}$ is associated with the height of the center of gravity of pedestrian, $d_L$ distance of launch and $t_{c}$ is the contact time. All of the variables are in the International System of Units (SI).

As a crucial result, the modeling permits estimating the vehicle's initial speed at the collision segment in terms of some variables collected by the investigators at the scene, as shown in Eq. \ref{Vva-1}. This velocity is critical to reconstructing the accident, and it has significant implications for the legal consequences.

\begin{equation}
V_{va}= \frac{(m_{v}+m_{p})\frac{ (g \mu_{p} \cos \beta \pm \sin \beta )}{ (\cos \theta + \mu_{p} \sin^2 \theta)} \left(\sqrt{\frac{t_{c}^2}{4} - \frac{2(\cos \theta + \mu_{p} \sin^2 \theta)}{g(\mu_{p} \cos \beta \pm \sin \beta)}(h_{cg}\mu_{p}-h_{cg} \tan \beta -d_{L})}-\frac{t_{c}}{2}\right)}{\alpha m_{v}}
\label{Vva-1}
\end{equation}

Eq. \ref{Vva-1} is obtained from Eq. \ref{Vva}. Likewise, it is known the vehicle initial speed at post-collision segment considering that $V_{pd} =V_{vd}$ ($\alpha=1$). $m_{v}$ and $m_{p}$ are associated to the mass of vehicle and the pedestrian, respectively.

It is crucial to bear in mind that for collisions with low frontal geometry vehicles, the launch angle of pedestrian es $\theta \neq 0$. On the contrary, for vehicles with high frontal geometry, the pedestrian's launch angle is  $\theta = 0$.
\\
\\
Regarding the accident avoidability study, it is necessary to consider the physics in the pre-collision segment. Here the movement of the car is uniform. Thus, the distance traveled from the perception point until the reaction point is related to perception-reaction time $t_{pr}$  with a constant speed $V_{vi}$. The Eq. \ref{Avoid} shows the relation of the kinematic variables.

\begin{equation}
V_{vi}=\frac{d_{pr}}{t_{pr}}
\label{Avoid}
\end{equation}
\\
\\
The following section, \ref{results}, shows the model's validation using experimental cases reported with dummies and bodies. The results gave the foundations to developing and implementing an affordable and handy mobile app to support the traffic police agencies and the experts during its investigation work.

\section{Results}
\label{results}

The model proposed in the previous section \ref{model} permits estimating the initial collision speed $V_{va}$  for a frontal collision vehicle-pedestrian as a function of the pedestrian's launch distance $d_{L}$. The model's validation develops using experimental controlled tests reported previously and data sets collected from reconstructed related cases. Two groups of data were selected as follows.
\\
The first data set corresponds to empirical values of $V_{va}$ and $d_{L}$ registered on \cite{Lenz1982}. Here, the authors collected reconstructed cases by experts related to frontal collisions and vehicle (low geometry)-pedestrian (adult). Furthermore, this work reports data corroborated with experimental tests with dummies (adult size) and bodies.
The second set of empirical data reported on \cite{randles01,dett97,sturtz76,severy66} are related to a vehicle's experimental tests (high geometry)-pedestrian frontal collisions. It was performed with child-sized dummies. \\
The experimental data sets were constituted by fifty (68) cases of vehicle-pedestrian collision  as follows:
Fifty-eight (58) cases of a vehicle (low geometry) with adult-sized constitued by  Dummies (22), adult bodies (18), and scene reconstructions (18). For another hand, Ten (10) vehicles (high geometry) with child-sized dummies. The registered variables are presented in the Appendix Section \ref{anexo}.    

The input values for the model validation are presented in Table \ref{inputvalues} and described below. According to the mechanism explained in section \ref{MD} input features are described:

\begin{itemize}
 
 \item For vehicle (low geometry)-pedestrian collision, the contact time between the first and second impact was taken as $t_{c}$=0.1s and the ratio of the speeds $\alpha$=1. 
    
    \item For vehicle (high geometry)-pedestrian collision, the contact time $t_{c}$ and the launch angle $\theta$ are equal to zero.
    
    \item In the case of the dummies-sized child, the center of gravity's exact values were unknown. Here a typical value was taken corresponding to  $h_{cg}$= 0.40 m, taking the child height as $h$=0.80 m.
    
    \item All of the experimental tests were on asphalt surface without tilt, which means $\beta$=0. 
    
    \item The mass of dummies sized-adult and bodies were given by the experiments used for model validation. For another hand, the mass of dummies sized-child was unknown; hence, it was assigned a typical value of $m_{p}$=30 Kg..
   
     \item The mass of vehicles $m_{v}$ were ranged between 750 Kg to 1474 Kg. Whereas  $m_{v}$  values for collisions with dummies sized-child were unknown;  Therefore, a typical value was assigned as 1200 Kg. Additionally, the mass of drivers was assumed with a nominal weight of 75Kg. The driver's mass is added to $m_{v}$ to obtain the total mass. 

\end{itemize}

\begin{table}[htp]
    \centering
    \begin{tabular}{|c|l|c|}\hline
        Notation & Variable & Value\\\hline
        $\alpha$ & \multicolumn{1}{p{8.5cm}|}{Ratio of the pedestrian and the vehicle speeds $\alpha=V_{pd}/V_{vd}$} & 1.0\\\hline
        $\beta$ & Tilt of the road & $0^o$\\\hline
        $\theta$ & \multicolumn{1}{p{5.0cm}|}{Launch angle of pedestrian (high geometry of vehicle)} & $0^o$\\\hline
        $\mu_p$ & Drag coefficient of pedestrian  & $0.7\leq\mu_p\leq 0.8$\\\hline
        $g$ & Gravity acceleration & $9.8m/s^2$\\\hline
        $h_p$ & Height of pedestrian (adult) & $1.55\leq h_p\leq 1.84$ \\\hline
        $h_{cg}$ & \multicolumn{1}{p{8.5cm}|}{Height of pedestrian (adult) center of gravity, which corresponds to half the weight (h)} & $h_{cg}=h_p/2$ \\\hline
        $t_c$ & Contact time (low geometry vehicle) & $\geq 0.1 s$\\\hline
        $t_c$ & Contact time (high geometry vehicle) & $ 0 s$\\\hline
        $m_v$ & Mass of vehicle & $750 kg\leq m_v\leq 1474kg$  \\\hline
        $m_{pa}$ & Mass of pedestrian (adult) & $60kg\leq m_{pa}\leq 89kg$\\\hline
        $m_{pn}$ & Mass of pedestrian child & $30kg$\\\hline
    \end{tabular}
    \caption{Input values for model validation}
    \label{inputvalues}
\end{table}

\begin{figure}[h]
\begin{center}
\includegraphics[scale=0.5]{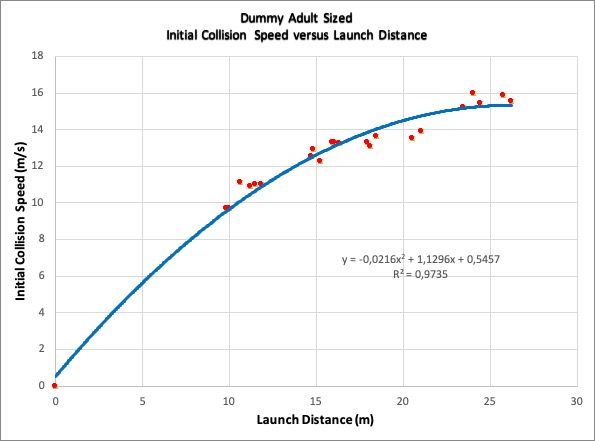}
\caption{Initial Collision Speed versus Launch Distance of Pedestrian for model and experiment with Dummy adult-sized.}
\label{validation}
\end{center}
\end{figure}

The model validation was developed by comparing the values obtained for the initial collision speed $V_{va}$ for the same launch distance of pedestrian  $d_{L}$ in each experimental case reported. The results are exhibit in Table \ref{modelvali}. Here, the $V_{va}$ Exp is the initial collision speed for the experimental case, and the $V_{va}$ Model is the one given by modeling.
Firstly, it was made a graphical comparison as depicted in Fig. \ref{validation}. This figure represents the values (red points) and its polynomial regression (blue curve) using the data experimental and calculated data for Dummy adult-sized. Due to the values being too close, there is an overlapping. It is difficult to distinguish between them; It happened for all kinds of experiments considered in this work, such as bodies, reconstructed cases, and dummies child-sized.  Therefore it was necessary a statistical comparison through variance analysis. It was made to guarantee the validity of the results.

\begin{table}[htp]
    \centering
    \begin{tabular}{|c|c|c||c|c|c|}
    \hline
    \multicolumn{3}{|c||}{Dummy adult-sized} & \multicolumn{3}{|c|}{Adult bodies}\\
    \hline
       $dL(m)$ & $V_{va}(m/s)$ Exp & $V_{va}(m/s)$ Model & $dL(m)$ & $V_{va}(m/s)$ Exp & $V_{va}(m/s)$ Model \\
    \hline
        10 & 9.69 & 9.69 & 8.5 & 9.58 & 9.58\\
    \hline
        9.8 & 9.72 & 9.72 & 15 & 11.89 & 11.89\\
    \hline
    11.5 & 11 & 11 & 10.2 & 11.22 & 11.22\\
    \hline
    11.2 & 10.88 & 10.88 & 11.5 & 11.14 & 11.14\\
    \hline
    11.8 & 11.02 & 11.02 & 11.1 & 11.1 & 11.1\\
    \hline
    14.7 & 12.56 & 12.56 & 9.8 & 11.24 & 11.47\\
    \hline
    10.6 & 11.14 & 11.14 & 9.2 & 11.01 & 11.17\\
    \hline
    15.2 & 12.25 & 12
    25 & 8.8 & 11.1 & 11.33\\
    \hline
    17.9 & 13.31 & 13.31 & 22.5 & 13.7 & 13.25\\
    \hline
    16 & 13.28 & 13.28 & 17.4 & 13.04 & 13.06\\
    \hline
    15.9 & 13.33 & 13.33 & 12.8 & 12.53 & 12.53\\
    \hline
    20.5 & 13.5 & 13.5 & 12.2 & 10.29 & 8.83\\
    \hline
    21 & 13.89 & 13.89 & 11.9 & 10.26 & 9.06\\
    \hline
    18.4 & 13.64 & 13.64 & 26.6 & 15.56 & 15.56\\
    \hline
    18.1 & 13.11 & 13.11 & 29.5 & 15.96 & 15.42\\
    \hline
    16.3 & 13.25 & 13.25 & 25.5 & 15.44 & 15.44\\
    \hline
    14.8 & 12.94 & 12.94 & 23.6 & 15.69 & 15.69\\
    \hline
    23.4 & 15.22 & 15.22 & 25.6 & 15.67 & 15.67\\
    \hline
    25.7 & 15.89 & 15.89 & 0 & 0 & 0\\
    \hline
    24.4 & 15.44 & 15.44 & 0 & 0 & 0\\
    \hline
    26.2 & 15.53 & 15.53 & 0 & 0 & 0\\
    \hline
    24 & 15.97 & 15.97 & 0 & 0 & 0\\
    \hline
    \end{tabular}
    \caption{Model validation}
    \label{modelvali}
\end{table}

\begin{table}[htp]
	\begin{center}
		\scalebox{1.1}{
		\begin{tabular}{|c|c|c|}
				\hline 
				Cases & $F_{value}$ & $F_{Critical}$ \\ 
				\hline 
				Dummy adult-sized & N/A & N/A  \\ \hline 
			Adult bodies & 0,76  & 4,45  \\ \hline 
				Scene Reconstruction & 0,42 &  5,31   \\ 	\hline 
		Dummy child sized & 0,05 & 5,11 \\
		\hline
			\end{tabular} }
		\caption{Variance Analysis. $F_{value}$ Calculated and $F_{critical}$ to compare the experimental and the model data for each case. }
		\label{ANOVA}
	\end{center}
\end{table}

\subsection{Variance Analysis}
The variance analysis ANOVA permits determine the variance between methods \cite{Dios10}. Here, it corresponds to the comparison of the initial collision speed $V_{va}$  from experiments and the model.
ANOVA uses the Fisher Test to verify two hypotheses, a null and an alternative. The null means that the data groups do not have significant differences, and the alternative one is the contrary. This method uses the indicator  F$_{critical}$ obtained from a Fisher distribution with a level of significance $5 \%$. When $F_{value}$ from data is calculated, and it is larger than the $F_{critical}$ the null hypothesis is rejected. Table \ref{ANOVA} shows the $F_{value}$ calculated and its $F_{critical}$ for Adult bodies, Scene Reconstruction and Dummy child-sized  cases.  It is important to note that the first row is related to dummies adult-sized; here, there is a N/A label, which means it does not apply; the results were the same for the initial speed values with four significant figures.
The results presented that for all cases, $F_{value}$ is less than $F_{critical}$, which means that the Initial Collision Speed from the Modelling does not have a significant statistical difference in comparison with the experimental cases.  The validation gave a very satisfactory result demonstrating the reliability of the model.

\newpage

\section{Conclusions and Discussion}
This paper presented a handle Vehicle-Pedestrian collision model used to build a free Mobile App called ARTgrama \cite{ezequielApp} available on google store to support the traffic accident investigators in their fieldwork and investigations.  
\\
\\
Modeling included detailed traffic accident analysis by segments called pre-collision, collision, and post-collision. The Pre-collision describes the driver perception and reaction phases. On the other hand, post-collision includes contact, flight, and sliding stages of the pedestrian as depicted in Fig \ref{col}. The model permits estimated the vehicle's initial speed, namely the velocity just before the collision as a function of pedestrian launch distance. It is also possible to identify the impact area on the road using the final positions of the center of gravity of the pedestrian and the front of the vehicle; It is crucial when the environmental conditions or the accident features do not allow located it. Additionally,  the traffic accident model by segments permits obtains a quantitative physical description for the pre-collision moment. This approach could be beneficial to road safety studies through the accident avoidability analysis.
\\
\\
The model validation was carried out by the comparison with sixty-eight (68) experimental cases reported in previous works, fifty (50) of them designed in the Laboratory using dummies (adult and child) and eighteen (18) data from real cases. A graphic and variance analysis demonstrated no significant statistical differences between the vehicle's initial collision speed value from modeling and the experiments. Therefore, it revealed that the model is suitable and reliable for the vehicle-pedestrian accident reconstruction for frontal collision with high and low vehicle geometry. It is also useful for Vehicle-cyclist and Vehicle-motorcyclist if it's the height of the center of gravity is known.
\\
\\
For reconstruction supported by Artgrama App, it is essential some data inputs obtained at the accident scene. Therefore, the app works as an investigation tool that provides a checklist to optimize the fieldwork, which means collecting the proper physical evidence, for instance, the vehicle's final position and the pedestrian, the pool of blood location, among the others, which are valuable for the reconstruction and the investigation of the accident.
\\
\\
Moreover, the Artgramma app is available in English and Spanish, allowing experts from middle and low-income countries to use it. It is valuable because there is no computer software access with high cost and training in their language in some regions.  Furthermore, most of the traffic accident investigators from isolated places are not specialized enough in physics and engineering fundamentals the reconstruction. Hence, Artgramma may contribute to the law enforcement investigators and experts in traffic to doing appropriate fieldwork and reconstruction of accidents that involve vulnerable road users as pedestrian, cyclist, and motorcycles.  Consequently,  better data quality about traffic accidents' causes and dynamics may improve road safety policies and strategies in many countries.

\section{Acknowledgments}
Universidad de Baja California is acknowledged who promotes the development of this project. Thanks to Defensor\'ia del Pueblo de Colombia, who permitted the validation of the model and the ArtGramma Mobile App for judicial cases. Likewise, the Escuela de Seguridad Vial of Colombia Police and Universidad Libre (Cali) are acknowledged for this project's academic application. Finally to the Universidad Antonio Nari\~no is thanked for allowing authors the time to undertake the study

\section{Declaration of Interest Statement}
It is stated that there are no significant competing interests, including financial or non-financial, professional, or personal interests interfering with the complete and objective presentation of the work described in this manuscript.



\newpage



\newpage

\bibliographystyle{elsarticle-num-names}
\bibliography{Arxiv}






\newpage
\section{Appendix}
\label{anexo}
\begin{table}[h]
    \centering
    \begin{tabular}{|c|c|c|c|c|c|c|}
    \hline
        $N^o$ & $dL(m)$ & $V_{va}(m/s)$ & $m_v$ $(kg)$ & $m_p$ $(kg)$ & $h_p$ $(m)$ & Vehicle model \\
        \hline
        1 & 10 & 9.69 & 750 & 70 & 1.65 & VOLKSWAGEN 1302 \\
        \hline
        2 & 9.8 & 9.72 & 750 & 70 & 1.65 & VOLKSWAGEN 1302 \\
        \hline
        3 & 11.5 & 11 & 865 & 74 & 1.62 & PEUGEOT 204  \\
        \hline
        4 & 11.2 & 10.88 & 865 & 74 & 1.62 & PEUGEOT 204 \\
        \hline
        5 & 11.8 & 11.02 & 865 &  74 & 1.62 & PEUGEOT 204 \\
        \hline
        6 & 14.7 & 12.56 & 865 & 74 & 1.62 & PEUGEOT 204 \\
        \hline
        7 & 10.6 & 11.14 & 650 &  65 & 1.64 & RENAULT 4  \\
        \hline
        8 & 15.2 & 12.25 & 1474 & 78 & 1.63 & MERCEDES 230 \\
        \hline
        9 & 17.9 & 13.31 & 1474 & 78 & 1.63 & MERCEDES 230 \\
        \hline
        10 & 16 & 13.28 & 1474 & 78 & 1.63 & MERCEDES 230 \\
        \hline
        11 & 15.9 & 13.33 & 1474 & 78 & 1.63 & MERCEDES 230 \\
        \hline
        12 & 20.5 & 13.5 & 1110 & 59 & 1.73 & AUDI 100 \\
        \hline
        13 & 21 & 13.89 & 1110 & 59 & 1.73 & AUDI 100 \\
        \hline
        14 & 18.4 & 13.64 & 1110 & 59 & 1.73 & AUDI 100 \\
        \hline
        15 & 18.1 & 13.11 & 925 & 70 & 1.64 & CITROEN GS \\
        \hline
        16 & 16.3 & 13.25 & 925 & 70 & 1.64 & CITROEN GS \\
        \hline
        17 & 14.8 & 12.94 & 925 & 70 & 1.64 & CITROEN GS \\
        \hline
        18 & 23.4 & 15.22 & 965 & 70 & 1.7 & MORIS MARINA \\
        \hline
        19 & 25.7 & 15.89 & 965 & 70 & 1.7 & MORIS MARINA \\
        \hline
        20 & 24.4 & 15.44 & 965 & 70 & 1.7 & MORIS MARINA \\
        \hline
        21 & 26.2 & 15.53 & 965 & 70 & 1.7 & MORIS MARINA \\
        \hline
        22 & 24 & 15.97 & 965 & 70 & 1.7 & MORIS MARINA \\
        \hline
    \end{tabular}
    \caption{Dummy adult-sized Experiments}
    \label{dummyexp}
\end{table}

\begin{table}[h]
    \centering
    \begin{tabular}{|c|c|c|c|c|c|c|}
    \hline
        $N^o$ & $dL(m)$ & $V_{va}(m/s)$ & $m_v$ $(kg)$ & $m_p$ $(kg)$ & $h_p$ $(m)$ & Vehicle model \\
        \hline
        1 & 8.5 & 9.58 & 750 & 73 & 1.58 & VOLKSWAGEN 1302 \\
        \hline
        2 & 15 & 11.89 & 865 & 53 & 1.63 & PEUGEOT 204 \\
        \hline
        3 & 10.2 & 12.22 & 865 & 78 & 1.64 & PEUGEOT 204  \\
        \hline
        4 & 11.5 & 11.14 & 865 & 85 & 1.69 & PEUGEOT 204 \\
        \hline
        5 & 11.1 & 11.1 & 865 &  75 & 1.72 & PEUGEOT 204 \\
        \hline
        6 & 9.8 & 11.47 & 650 & 81 & 1.75 & RENAULT 4  \\
        \hline
        7 & 9.2 & 11.17 & 650 &  74 & 1.79 & RENAULT 4  \\
        \hline
        8 & 8.8 & 11.33 & 650 & 89 & 1.84 & RENAULT 4 \\
        \hline
        9 & 22.5 & 13.25 & 1474 & 43 & 1.55 & MERCEDES 230 \\
        \hline
        10 & 17.4 & 13.06 & 1474 & 53 & 1.66 & MERCEDES 230 \\
        \hline
        11 & 12.8 & 12.53 & 1474 & 78 & 1.68 & MERCEDES 230 \\
        \hline
        12 & 12.2 & 8.83 & 750 & 72 & 1.74 & VOLKSWAGEN GOLF \\
        \hline
        13 & 11.9 & 9.06 & 750 & 83 & 1.83 & VOLKSWAGEN GOLF \\
        \hline
        14 & 26.6 & 15.56 & 965 & 70 & 1.7 & MORIS MARINA \\
        \hline
        15 & 29.5 & 15.42 & 965 & 70 & 1.7 & MORIS MARINA \\
        \hline
        16 & 25.5 & 15.44 & 965 & 70 & 1.7 & MORIS MARINA \\
        \hline
        17 & 23.6 & 15.69 & 965 & 70 & 1.7 & MORIS MARINA \\
        \hline
        18 & 25.6 & 15.67 & 965 & 70 & 1.7 & MORIS MARINA \\
        \hline
    \end{tabular}
    \caption{Adult bodies experiments}
    \label{adultbodexp}
\end{table}

\begin{table}[h]
    \centering
    \begin{tabular}{|c|c|c|c|c|c|c|}
    \hline
        $N^o$ & $dL(m)$ & $V_{va}(m/s)$ & $m_v$ $(kg)$ & $m_p$ $(kg)$ & $h_p$ $(m)$ & Vehicle model \\
        \hline
        1 & 8.2 & 9.72 & 750 & 70 & 1.65 & VOLKSWAGEN 1302 \\
        \hline
        2 & 14.5 & 11.1 & 865 & 74 & 1.62 & PEUGEOT 204 \\
        \hline
        3 & 14.5 & 11.1 & 865 & 74 & 1.62 & PEUGEOT 204 \\
        \hline
        4 & 14.5 & 11.1 & 865 & 74 & 1.62 & PEUGEOT 204 \\
        \hline
        5 & 14.5 & 11.1 & 865 & 74 & 1.62 & PEUGEOT 204 \\
        \hline
        6 & 17 & 12.5 & 650 & 65 & 1.64 & RENAULT 4  \\
        \hline
        7 & 17 & 12.5 & 650 & 65 & 1.64 & RENAULT 4  \\
        \hline
        8 & 17 & 12.5 & 650 & 65 & 1.64 & RENAULT 4  \\
        \hline
        9 & 16 & 12.5 & 1474 & 78 & 1.63 & MERCEDES 230 \\
        \hline
        10 & 16 & 12.5 & 1474 & 78 & 1.63 & MERCEDES 230 \\
        \hline
        11 & 16 & 12.5 & 1474 & 78 & 1.63 & MERCEDES 230 \\
        \hline
        12 & 28 & 13.88 & 1110 & 59 & 1.73 & AUDI 100 \\
        \hline
        13 & 28 & 13.88 & 1110 & 59 & 1.73 & AUDI 100 \\
        \hline
        14 & 28 & 13.88 & 1110 & 59 & 1.73 & AUDI 100 \\
        \hline
        15 & 9.1 & 8.8 & 750 & 75 & 1.74 & VOLKSWAGEN GOLF \\
        \hline
        16 & 7 & 13.3 & 965 & 70 & 1.64 & CITROEN GS \\
        \hline
        17 & 20.1 & 16.4 & 965 & 70 & 1.7 & MORIS MARINA \\
        \hline
        18 & 29 & 15.3 & 965 & 60 & 1.7 & MORIS MARINA \\
        \hline
    \end{tabular}
    \caption{Scenes reconstructions (vehicle-adult pedestrian collision)}
    \label{scencesrec}
\end{table}

\begin{table}[h]
    \centering
    \begin{tabular}{|c|c|c|c|c|c|}
    \hline
        $N^o$ & $dL(m)$ & $V_{va}(m/s)$ & $m_v$ $(kg)$ & $m_p$ $(kg)$ & Pedestrian height $(m)$ \\
        \hline
        1 & 7.9 & 9.16 & 1200 & 30 & 0.8 \\
        \hline
        2 & 13 & 11.11 & 1200 & 30 & 0.8 \\
        \hline
        3 & 15.3 & 12.22 & 1200 & 30 & 0.8 \\
        \hline
        4 & 39 & 20.55 & 1200 & 30 & 0.8 \\
        \hline
        5 & 13.8 & 13.88 & 1200 & 30 & 0.8 \\
        \hline
        6 & 27.8 & 20.27 & 1200 & 30 & 0.8 \\
        \hline
        7 & 43.9 & 21.94 & 1200 & 30 & 0.8 \\
        \hline
        8 & 18 & 15.55 & 1200 & 30 & 0.8 \\
        \hline
        9 & 40.8 & 17.7 & 1200 & 30 & 0.8 \\
        \hline
        10 & 34 & 17.7 & 1200 & 30 & 0.8 \\
        \hline
    \end{tabular}
    \caption{Dummy child-sized experiments}
    \label{dummychildexp}
\end{table}

\end{document}